\documentclass[traditabstract]{aa}
\usepackage     [utf8]                 {inputenc}  
\usepackage{verbatim} 
\usepackage{amsmath}
\usepackage{amssymb}
\usepackage{graphicx}
\usepackage{flafter} 
\usepackage{booktabs}
\usepackage[authoryear]{natbib}
\usepackage[colorlinks=true,linkcolor=blue,urlcolor=blue,citecolor=blue]{hyperref}
\usepackage{xspace}
\usepackage{siunitx}
\usepackage{lscape}
\usepackage{longtable}
\usepackage{threeparttable}

\makeatletter

\usepackage{txfonts}\usepackage{twoopt}
\bibpunct{(}{)}{;}{a}{}{,}
\usepackage{enumitem}
\newcommand{\ca}{\mbox{Ca\,{\textsc{ii}}~K\,}}

\usepackage[rightcaption]{sidecap}
\graphicspath{ {images/} }

\titlerunning{Solar Irradiance Reconstruction over the Telescopic Era Using a Revised Photospheric Magnetic Field Model}
\authorrunning{D.~Temaj}

\makeatother

\begin{document}

\title{Solar Irradiance Reconstruction over the Telescopic Era Using a Revised Photospheric Magnetic Field Model \\}
\author{D.~Temaj\inst{\ref{MPS},\ref{TUBS}}\thanks{temaj@mps.mpg.de}, N.A.~Krivova\inst{\ref{MPS}}, 
T.~Chatzistergos\inst{\ref{MPS}},
S.K.~Solanki\inst{\ref{MPS}}, 
B.~Hofer\inst{\ref{MPS}}
}

\institute{
Max-Planck-Institut für Sonnensystemforschung, Justus-von-Liebig-Weg 3, Göttingen, Germany\label{MPS}
\and Technische Universität Braunschweig, Universitätsplatz 2, 38106, Braunschweig, Germany\label{TUBS}
}

\date{\today}

\abstract{
The Sun is the primary source of energy for Earth and one of the main external drivers of its climate. 
Solar irradiance~-- the radiative power emitted by the Sun and received at 1~AU~-- varies on all observable timescales.
It is measured as total solar irradiance (TSI), the spectrally integrated flux, or as spectral solar irradiance (SSI), its wavelength-dependent distribution.
However, direct space-based irradiance measurements span only about five decades and are too short to capture long-term trends, making reconstructions crucial for studying solar influence on climate. On climate-relevant timescales, irradiance variations are driven by changes in the solar surface magnetic field, which form the basis of reconstructions guided by physics. 

Here we present revised reconstructions of TSI and SSI over the past four centuries using the physics-based SATIRE-T (Spectral And Total Irradiance REconstruction, for the Telescopic era) model. 
SATIRE-T relates irradiance variability to the evolution of the solar surface magnetic field inferred from sunspot number records. 
In this work, we implement a recently revised description of magnetic field evolution that more realistically links the emergence of small-scale magnetic features to sunspot activity, constrained by modern observations.
Using two independent sunspot number series as input, we obtain consistent reconstructions of magnetic flux and solar irradiance.
The model reproduces the observed or independently reconstructed total and open magnetic flux, and agrees closely with satellite measurements of TSI and Lyman-$\alpha$ irradiance, with correlation coefficients of 0.81--0.98 for 81-day-smoothed space-based TSI records, 0.69--0.85 for TSI at daily cadence, and 0.92 for daily Lyman-$\alpha$ irradiance.
On secular timescales, the reconstructed TSI increases by 0.67--0.75\,W/m$^2$ between the 50-year means over 1650--1700 and 1967--2017.
}


\maketitle

\section{Introduction}

The Sun provides most of the energy to Earth \citep[e.g.][]{kren_where_2017}, sustaining life and affecting climate. The radiative flux or radiative power received per unit area at a distance of 1\,AU is termed solar irradiance. 
The spectrally integrated flux is known as the
Total Solar Irradiance (TSI), while the wavelength-dependent quantity is the Spectral Solar Irradiance (SSI).
Solar irradiance has been measured from space since 1978, revealing variability on all observable timescales, from minutes to decades and longer \citep[e.g.,][]{willson_solar_1988,frohlich_solar_2006, kopp_magnitudes_2016,kopp_solar_2025}.
However, the instrumental record is too short to compare directly with climate data and to assess the Sun's impact on the Earth's climate. 
Reconstructions are therefore required to extend irradiance records back in time. 

Solar irradiance variability on timescales longer than a few hours is driven by changes in solar surface magnetism, specifically, the darkening caused by sunspots and the brightening due to faculae and network regions \citep[e.g.,][]{krivova_reconstruction_2003, yeo_solar_2017, shapiro_nature_2017}. 
This means that to model irradiance variations, knowledge of the evolving solar surface magnetic field is required. 
Direct magnetic field measurements have been available only for the past few decades \citep[see, e.g.,][]{pevtsov_long-term_2021}.
Thus, to go further back in time, proxies of solar magnetic activity are needed, particularly those that distinguish between sunspots and faculae.

Sunspot numbers (SN) provide the longest direct observational record of solar activity, spanning more than 400~years \citep[see reviews by][]{arlt_historical_2020,clette_recalibration_2023}.
However, direct information on bright features, such as faculae and the network, is missing.
Historical full-disc photographs of the Sun in \ca line, available since the late 19th century \citep{chatzistergos_full-disc_2022}, are an excellent source of such information \citep{chatzistergos_analysis_2020}, as they capture the chromospheric counterparts of faculae.
While significant progress has been made in recovering and processing these archives \citep{chatzistergos_analysis_2018,chatzistergos_full-disc_2022}, further work is required
before they can reliably be used for long-term irradiance reconstructions \citep{chatzistergos_understanding_2024}.
Consequently, for periods lacking direct facular observations, most existing long-term reconstructions have inferred the facular contribution indirectly from sunspot or cosmogenic-isotope data under various, often uncertain, assumptions.
This leads to significant differences in the reconstructed long-term trends of solar irradiance ranging from near-zero secular change to several W\,m$^{-2}$ \citep[for a review of various models and resulting uncertainties, see][]{chatzistergos_long-term_2023}.
Particularly problematic are extended quiet periods, such as grand minima of solar activity, when only a few to no sunspots emerge over years or even decades.
Over such periods, no information on the emergence of the magnetic field on the Sun is available, resulting in underestimated magnetic flux and overestimated secular irradiance variability \citep[for a discussion, see][]{krivova_modelling_2021}.

To address this, \cite{krivova_modelling_2021} revised the magnetic flux evolution model originally developed by \cite{solanki_evolution_2000, solanki_secular_2002}, to incorporate a more realistic representation of small-scale magnetic features such as faculae and network.
The updated model relies on modern observations showing that the emergence rates of magnetic regions follow a single power-law distribution across a wide range of fluxes \citep{thornton_small-scale_2011}. 
This approach gives a more realistic reconstruction of the solar magnetic flux and improved agreement with independent observations and reconstructions.

Among existing irradiance models \citep[for a review, see, e.g.,][]{chatzistergos_long-term_2023}, the Spectral And Total Irradiance REconstructions (SATIRE) framework provides a robust, semi-empirical, physics-based approach \citep{krivova_towards_2011,solanki_solar_2013}. 
Different SATIRE versions use different observational data as input. 
SATIRE-S (for the satellite era; \citealt{krivova_reconstruction_2003,yeo_reconstruction_2014,chatzistergos_revisiting_2025}) employs full-disc magnetograms and continuum images, providing the highest accuracy but limited temporal coverage (roughly 50~years).
SATIRE-T (for the telescopic era; \citealt{krivova_reconstruction_2007,krivova_reconstruction_2010,wu_solar_2018}) extends the reconstruction back to the Maunder minimum by using sunspot numbers to model the evolution of the solar surface magnetic field and subsequently the irradiance variability.
 
In this work, we employ the SATIRE-T model \citep[][]{krivova_reconstruction_2007, krivova_reconstruction_2010, wu_solar_2018} together with the revised magnetic flux model of \cite{krivova_modelling_2021} to reconstruct solar surface magnetic field evolution and solar irradiance over the entire telescopic era, extending back to the Maunder minimum. 
The paper is structured as follows: Sect.~\ref {sect: Data} describes the data used, Sect.~\ref {sect: methods} outlines the modelling approach, Sect.~\ref {Sect: results} presents and discusses the results, and Sect.~\ref{conclusion} summarises the main findings.

\section{Data}\label{sect: Data}
\subsection{Input data}

To describe the emergence of the solar surface magnetic field (see Sect.~\ref{sect: mf_model}), we use two independent sunspot number (SN) records as input: the International Sunspot Number version 2.0 \citep[ISNv2;][]{clette_recalibration_2023}\footnote{\url{https://www.sidc.be/SILSO/datafiles}} and the Group Sunspot Number (GSN) series by \citet[][hereafter CEA17]{chatzistergos_new_2017}\footnote{\label{sunclimatewebsite}\url{https://www2.mps.mpg.de/projects/sun-climate/data.html}}. 
ISNv2 provides daily values since 1818, monthly data from 1749, and annual values from 1700. 
Here we combine daily values from 1818 onward with monthly ones back to 1749, interpolated to daily resolution.
The CEA17 series begins in 1739, but data prior to 1749 are sparse and irregular and are therefore excluded.
This record ends in 2010, since raw group counts in the \cite{vaquero_revised_2016} database are not available afterwards.
We linearly regressed the CEA17 record against ISNv2 over their 1976--2010 overlap to match their scales and extended CEA17 to the present using ISNv2.

To extend these records back to the beginning of telescopic observations, we also incorporate the GSN series by \citet[][hereafter HoSc98]{hoyt_group_1998},
which provides daily data since 1610.
However, because the earliest observations contain substantial gaps, we use the record only after 1650. 
We linearly scaled the HoSc98 series to separately match ISNv2 and CEA17, respectively, and used the scaled versions to extend these records from 1650 to 1749.
As an alternative extension, we also used the annual sunspot numbers estimated with the active day fraction method by \citet{carrasco_relationship_2022,carrasco_understanding_2024,carrasco_numerical_2025}.
This record covers 1635--1721 and 1725--1726, but we excluded the latter period due to discontinuity with the earlier record.
We used this series to extend ISNv2 back to 1635. 
The gap between 1721 and 1749 in this case was filled by interpolating annual ISNv2 values to daily cadence.
Interpolation was also applied to fill all remaining gaps in the series, since the magnetic flux evolution model requires daily sunspot number input.

We note that the HoSc98 GSN has been criticised for potential inaccuracies \citep{clette_recalibration_2023,lockwood_assessment_2016, usoskin_solar_2016,chatzistergos_assessment_2025}, including misclassified zero-group days during the Maunder Minimum, that is, days with no available observations were assigned a SN group count of zero \citep{vaquero_revised_2016,clette_recalibration_2023}.
For this reason, we use it only where no other, more reliable data exist (i.e., before 1739) and treat it as a lower limit for solar activity during the Maunder Minimum. 
We emphasise, however, that the use of HoSc98 introduces increased uncertainty in our results for those early periods~-- a limitation shared by all studies extending over and prior to the Maunder minimum that rely on sunspot observations.
The series by \citet{carrasco_relationship_2022,carrasco_understanding_2024,carrasco_numerical_2025} provides a more realistic estimate of solar activity during that time, but its annual resolution limits its use for our purposes.
We therefore employ it to obtain a rough upper-limit estimate for our reconstruction during the Maunder Minimum.
ISNv2 values in the first decades of the 18th century may also be somewhat overestimated \citep[see e.g.][]{usoskin_solar_2021,carrasco_understanding_2024}.

To estimate the fractional area of the solar disc covered by sunspots (Sect.~\ref{sect: model irradiance}), we used the sunspot area record by \cite{mandal_sunspot_2020}, updated with recent observations\footref{sunclimatewebsite}. 
We extended this record backward to 1650 by fitting a second-degree polynomial to the corresponding sunspot number series and filled gaps through interpolation. 
After 2010, missing values were replaced with areas derived from SDO/HMI (Helioseismic and Magnetic Imager on board of Solar Dynamic Observatory) continuum observations \citep{chatzistergos_revisiting_2025}.

\subsection{Data used for the optimisation process}\label{sect: data optimisation}

As discussed in Sects.~\ref{sect: methods} and \ref{Sect: results}, model parameters were constrained by comparing model outputs to the following independent datasets: 
\begin{itemize}
     
\item Total magnetic flux from Carrington-rotation synoptic magnetograms obtained at the Wilcox Solar Observatory (WSO), Mount Wilson Observatory (MWO), and Kitt Peak National Solar Observatory (NSO/KP) \citep{arge_two_2002, wenzler_reconstruction_2006}. 
For the period 1976--2002, all three datasets are available and have been averaged; for 2002--2009, we averaged data from WSO and MWO.

\item Open magnetic flux from the empirical reconstruction based on the geomagnetic aa-index by \cite{lockwood_reconstruction_2024}, available for the period 1868--2022.8 as both Carrington-rotation (CR) and 13-rotation averages. For optimisation, we employ the 1-CR dataset. 

\item The TSI composite of measurements by \cite{montillet_data_2022}\footnote{
\url {ftp://ftp.pmodwrc.ch/pub/data/irradiance/virgo/TSI/TSI_composite/MergedPMOD_NobaselineScaleCycle23_JPM_July2025_v1.txt}}, covering the period 1980--2025 at daily cadence. For consistency, we scaled it to match the TSIS-1/TIM (Total and Spectral Solar Irradiance Sensor/ Total Irradiance Monitor) absolute level at the minimum of cycle~24 (2019.96$\pm$0.5~yr). 
\end{itemize}

\subsection{Data used for validation and comparisons}\label{sect: data comarison}

To assess the performance of our model, we compare our OSF, TSI, and SSI reconstructions with several datasets not used in the optimization.

\paragraph{Open magnetic flux.}

\begin{itemize}
\item OSF reconstruction of \citet{wu_solar_2018}, derived from cosmogenic isotopes (thus independent of magnetograms and sunspot data).

\item In-situ OSF measurements of \citet[]{owens_sunward_2017}, for 1998--2018 at annual resolution, based on direct spacecraft measurements of the radial magnetic field.

\item The OSF estimates, also from \citet[]{owens_sunward_2017}, derived from the OMNI interplanetary magnetic-field data for 1964--1998 at annual resolution, corrected for solar-wind kinematic effects following \cite{lockwood_excess_2009}.
These provide an observationally based but proxy-corrected estimate of OSF.
For quantitative comparison, we use only the in-situ part of the \citet{owens_sunward_2017} series, but
the full series is still shown as an additional reference.

\end{itemize}

\paragraph{Total solar irradiance TSI.}

\begin{itemize}
\item The recent high-precision TSI measurements from TSIS-1/TIM\footnote{\url {https://lasp.colorado.edu/data/tsis/tsi_data/tsis_tsi_L3_c24h_latest.txt}} \citep[2018--2025.5;][]{pilewskie_tsis-1_2018}.

\item SORCE/TIM (Total Irradiance Monitor onboard the SOlar Radiation and Climate Experiment) data\footnote{\url {https://lasp.colorado.edu/data/sorce/tsi_data/daily/sorce_tsi_L3_c24h_latest.txt}} \citep[2003.15--2020.15][]{kopp_total_2005}. 

\item The TSI composite by \citet[][1978.9 to 2025.1]{dudok_de_wit_methodology_2017}\footnote{\url {https://spot.colorado.edu/~koppg/TSI/TSI_Composite-SIST.txt}}. 

\item TSI measurements from Variability of Irradiance and Gravity Oscillations \citep[VIRGO; 1996.14--2025.4;][]{frohlich_virgo_1995, finsterle_total_2021}\footnote{
\url {ftp://ftp.pmodwrc.ch/pub/data/irradiance/virgo/TSI/VIRGO_TSI_Daily_V8_20250801.txt}} instrument onboard the Solar and Heliospheric Observatory \citep[SoHO;][]{domingo_soho_1995}.

\item The TSI reconstruction of \cite{wu_solar_2018}, which is based on an earlier version of the SATIRE-T model.

\item The TSI reconstruction of \cite{chatzistergos_revisiting_2025}, which employs the SATIRE-S model. 
\end{itemize}

\paragraph{Spectral solar irradiance SSI.}
\begin{itemize}
 \item The Lyman-$\alpha$ composite of \cite{machol_improved_2019} which combines data from multiple instruments and fills data gaps using solar activity proxies: the $10.7\,\mathrm{cm}$ radio flux (1947--1955), the $30\,\mathrm{cm}$ radio flux (1955--1978), and the $\mathrm{Mg\,\textsc{ii}}$ index for the remaining gaps.
 \item The SSI reconstruction from \cite{chatzistergos_revisiting_2025}.
\end{itemize}

\section{Model}\label{sect: methods}

We use the SATIRE-T model to reconstruct solar irradiance variability. 
The model requires the distribution and the temporal evolution of magnetic features on the solar surface as input.
To obtain these, we first reconstruct the evolution of the solar surface magnetic flux using the model of the magnetic field evolution, originally proposed by \citet{solanki_evolution_2000}, subsequently refined by \citet{solanki_search_2002} and \citet{vieira_evolution_2010}, and most recently revised by \cite{krivova_modelling_2021}.
This latest revision explicitly accounts for the contribution of small-scale magnetic features emerging outside active regions, as described in Sect.~\ref{sect: mf_model}.
The reconstructed magnetic fluxes of the various components are then used as input to SATIRE-T to compute total and spectral solar irradiance. 
Although SATIRE-T has been described in detail in earlier papers \cite[e.g.,][]{krivova_reconstruction_2007,krivova_reconstruction_2010, wu_solar_2018}, we summarise its concept and key components in Sect.~\ref{sect: model irradiance} for completeness.

\subsection{Magnetic field evolution} \label{sect: mf_model}

To model the evolution of the solar magnetic field over the telescopic era, we used the model by \citet{krivova_modelling_2021}, which takes the sunspot number as input and describes the temporal evolution of the solar surface magnetic flux by solving a set of coupled ordinary differential equations. 
The model divides the entire range of the emerging magnetic regions into two classes: (1) active regions (ARs), containing sunspots, with magnetic fluxes above $4\times10^{20}\,\mathrm{Mx}$; and (2) small-scale emergences (SSEs), which do not contain sufficient flux to form sunspots. 
ARs include both sunspots and faculae.
SSEs include ephemeral regions (ERs, bipolar magnetic regions with typical fluxes of about $10^{18}\,\mathrm{Mx}$) and the yet smaller internetwork fields, with fluxes down to $10^{16}$~Mx (see \citealp{krivova_modelling_2021} for further details, as well as \citealt{van_driel-gesztelyi_evolution_2015}).
The number of the smallest internetwork elements ($10^{16}-10^{17}$~Mx) remains nearly constant over the solar cycle \citep{buehler_quiet_2013, lites_solar_2014}.

Frozen in the ionized solar wind plasma, part of the surface magnetic flux is carried outwards, forming the open magnetic flux $\phi_\mathrm{open}$. 
Following \cite{vieira_evolution_2010} and \citet{krivova_modelling_2021}, the model differentiates between the rapidly, $\phi_\mathrm{open}^\mathrm{r}$, and slowly, $\phi_\mathrm{open}^\mathrm{s}$, evolving components of the open flux. 
The rapid component arises mainly from small coronal holes associated with ARs, which form and disappear quickly on timescales of days to weeks.
The slowly evolving component is associated with large-scale polar coronal holes sustained by both ARs and SSEs. 
Its characteristic timescales range from months to years. 

The evolution of the various magnetic flux components in time, $t$, is described by the set of coupled differential equations: 
\begin{eqnarray}
\label{eq: total flux}
    \frac{\mathrm{d}\phi_\mathrm{AR}}{\mathrm{d}t}= \varepsilon_\mathrm{AR} - \frac{\phi_\mathrm{AR}}{\tau_\mathrm{AR}^\mathrm{0}} - \frac{\phi_\mathrm{AR}}{\tau_\mathrm{AR}^\mathrm{s}} - \frac{\phi_\mathrm{AR}}{\tau_\mathrm{AR}^\mathrm{r}},\\
    \frac{\mathrm{d}\phi_\mathrm{SSE}}{\mathrm{d}t}= \varepsilon_\mathrm{SSE} - \frac{\phi_\mathrm{SSE}}{\tau_\mathrm{SSE}^\mathrm{0}} - \frac{\phi_\mathrm{SSE}}{\tau_\mathrm{SSE}^\mathrm{s}},\\
    \frac{\mathrm{d}\phi_\mathrm{open}^\mathrm{r}}{\mathrm{d}t}= \frac{\phi_\mathrm{AR}}{\tau_\mathrm{AR}^\mathrm{r}} - \frac{\phi_\mathrm{open}^\mathrm{r}}{\tau_\mathrm{open}^\mathrm{r}},\\
    \frac{\mathrm{d}\phi_\mathrm{open}^\mathrm{s}}{\mathrm{d}t}= \frac{\phi_\mathrm{AR}}{\tau_\mathrm{AR}^\mathrm{s}} + \frac{\phi_\mathrm{SSE}}{\tau_\mathrm{SSE}^\mathrm{s}} - \frac{\phi_\mathrm{open}^\mathrm{s}}{\tau_\mathrm{open}^\mathrm{s}},\\[2mm]
\phi_\mathrm{open} =\phi_\mathrm{open}^\mathrm{r}  +\phi_\mathrm{open}^\mathrm{s},\\[2mm]
    \phi_\mathrm{total} =\phi_\mathrm{AR} +\phi_\mathrm{SSE} + \phi_\mathrm{open}.
\end{eqnarray}

Here, $\phi_\mathrm{AR}$ and $\phi_\mathrm{SSE}$ denote the magnetic flux in active regions and small-scale emergences, respectively. 
The first two equations describe the temporal balance between the emergence ($\varepsilon_\mathrm{AR}$ and $\varepsilon_\mathrm{SSE}$) and decay of magnetic flux in active regions and small-scale emergences, respectively.
The following two equations account for the transfer of magnetic flux from closed-field regions to the open flux, subdivided into the rapidly, $\phi_\mathrm{open}^\mathrm{r}$, and slowly, $\phi_\mathrm{open}^\mathrm{s}$, evolving components, whose sum gives the total open flux $\phi_\mathrm{open}$.
The sum of all components provides the total magnetic flux, $\phi_\mathrm{total}$, at the solar surface and open into the heliosphere.
The parameters $\tau_\mathrm{AR}^\mathrm{0}$ and $\tau_\mathrm{SSE}^\mathrm{0}$ denote the decay timescales of flux in ARs and SSEs, respectively, while $\tau_\mathrm{AR}^\mathrm{s}$, $\tau_\mathrm{AR}^\mathrm{r}$, and $\tau_\mathrm{SSE}^\mathrm{s}$ specify transfer timescales of the AR- and ER-flux to the open-flux components. 
Finally, $\tau_\mathrm{open}^\mathrm{r}$ and $\tau_\mathrm{open}^\mathrm{s}$ describe the decay of the rapid and slow components of the open flux, respectively.

Using observational constraints from \citet{parnell_power-law_2009} and \citet{thornton_small-scale_2011}, \citet{krivova_modelling_2021} estimated the typical lifetimes of magnetic flux in ARs and SSEs to be approximately 10 days for $\tau_\mathrm{AR}^\mathrm{0}$ and 6 minutes for $\tau_\mathrm{SSE}^\mathrm{0}$, respectively.
These values were derived by dividing the number of observed magnetic features (with fluxes above or below $\phi_\mathrm{AR}$ for AR and SSE regions, respectively) by their corresponding emergence rates.

The emergence rates of magnetic flux in ARs and SSEs, $\varepsilon_\mathrm{AR}$ and $\varepsilon_\mathrm{SSE}$, respectively, follow the observed flux-emergence distribution derived by
\citet{thornton_small-scale_2011}:
\begin{equation}
    \frac{\mathrm{d}N}{\mathrm{d}\phi}= \frac{n_\mathrm{0}}{\phi_\mathrm{0}} \left(\frac{\phi}{\phi_\mathrm{0}}\right)^m,
\label{Eq: Power law}
\end{equation}
where $\phi_\mathrm{0}= 10^{16}~\mathrm{Mx}$ is the flux in the smallest considered features, $n_0 = 3.14 \times 10^{-14}~\mathrm{cm^{-2} day^{-1}}$ is the total number density of emergence events, and $m = -2.69$ is the slope of the distribution.

However, flux emergence is not constant over the solar cycle.
\cite{harvey_properties_1993} showed that the emergence rate of large magnetic regions, with areas exceeding 3.5 square degrees (roughly corresponding to AR fluxes of $\ge 4\times10^{20}\, \mathrm{Mx}$; \citealt{hofer_influence_2024}) varied by a factor of $\sim 8.3$ between solar maximum and minimum of solar cycle 21. 
In contrast, smaller magnetic regions with areas 2.5–-3.5 square degrees (that is SSEs with fluxes of $\sim 3\times10^{18}-4\times10^{20}\,\mathrm{Mx}$) varied only by a factor of $\sim 2.1$. 
This behaviour is consistent with the observed quadratic relationship between sunspot and facular areas \citep[e.g.,][]{shapiro_variability_2014,yeo_how_2020, chatzistergos_scrutinising_2022}. 

To account for this solar cycle–dependent variability, \citet{krivova_modelling_2021} allowed the slope $m$ of the emergence-rate distribution to vary with activity (quantified by the SN):
\begin{equation}
    m(\mathrm{SN}) = m_\mathrm{1} - ( \mathrm{SN}_\mathrm{1}^\alpha - \mathrm{SN}^\alpha) \frac{ m_\mathrm{1}- m_\mathrm{2}}{\mathrm{SN}_\mathrm{1}^\alpha - \mathrm{SN}_\mathrm{2}^\alpha} , 
\end{equation}
where $\alpha$ is a free parameter constrained through comparison with observations. 

Following \cite{krivova_modelling_2021}, we adopt SN$_1\,{=}\,217$ and SN$_2\,{=}\,17$, corresponding to the maximum and minimum (average over the preceding and following minima) of cycle 21, consistent with the periods analysed by \citet{harvey_properties_1993}. 
To reproduce the observed \citep{harvey_properties_1993} ratios in emergence rates, the slope is allowed to vary symmetrically about the mean value $m$ by an amount $\Delta m = 0.0946$, yielding $m_\mathrm{1}=-2.59$ and $m_\mathrm{2} = -2.78$.

The emergence rates of ARs and SSEs are then computed by integrating Eq. (\ref{Eq: Power law}) over the respective flux ranges:
\begin{equation}
    \varepsilon_\mathrm{AR} = \int_{\phi_\mathrm{AR}}^{\phi_\mathrm{lim}} \frac{n_\mathrm{0}}{\phi_\mathrm{0}} \left(\frac{\phi}{\phi_\mathrm{0}}\right)^m \phi \, \mathrm{d}\phi =  \frac{n_\mathrm{0}}{(m+2) \phi_\mathrm{0}^{(m+1)}} \left( \phi_\mathrm{lim}^{(m+2)} -  \phi_\mathrm{AR}^{(m+2)}\right) , 
\end{equation}
\begin{equation}
    \varepsilon_\mathrm{SSE} = \int_{\phi_\mathrm{0}}^{\phi_\mathrm{AR}} \frac{n_\mathrm{0}}{\phi_\mathrm{0}} \left(\frac{\phi}{\phi_\mathrm{0}}\right)^m \phi \, \mathrm{d}\phi =  \frac{n_\mathrm{0}}{(m+2) \phi_\mathrm{0}^{(m+1)}} \left( \phi_\mathrm{AR}^{(m+2)} -  \phi_\mathrm{0}^{(m+2)}\right).
\end{equation} 
Following \citet{krivova_modelling_2021} and \citet{hofer_influence_2024}, we use $\phi_\mathrm{AR} = 4 \times 10^{20}~\mathrm{Mx}$, corresponding to the smallest regions capable of producing sunspots, and $\phi_\mathrm{lim}=10^{23}\,\mathrm{Mx}$ as the upper limit \cite[cf.][]{thornton_small-scale_2011}.
Regions above this limit are extremely rare, so the exact choice of $\phi_\mathrm{lim}$ has negligible influence \citep{hofer_influence_2024}.
We also recall that $\phi_\mathrm{0} = 10^{16}~\mathrm{Mx}$ corresponds to typical internetwork fluxes.

\subsection{Modeling irradiance variability}
\label{sect: model irradiance}

Variations in solar irradiance on time scales of days to millennia are caused by the evolution of the surface magnetic field through the competing effects of sunspot darkening and facular brightening. 
The SATIRE model computes irradiance as the sum of contributions from five surface components: sunspot umbrae (u), penumbra (p), faculae (f), network (n), and the quiet Sun (q).
With quiet Sun, we refer to surface areas free of (detectable) magnetic field.

Thus, solar irradiance at a given time and wavelength is computed as
\begin{equation}
\label{eq: TSI}
\begin{aligned}
F(\lambda, t)= & \alpha_{\mathrm{q}}(t) F_{\mathrm{q}}(\lambda)+\alpha_{\mathrm{u}}(t) F_{\mathrm{u}}(\lambda) \\
& +\alpha_{\mathrm{p}}(t) F_{\mathrm{p}}(\lambda)+\left[\alpha_{\mathrm{f}}(t)+\alpha_{\mathrm{n}}(t)\right] F_{\mathrm{f}}(\lambda),
\end{aligned}
\end{equation}
where $\alpha_i$ are the fractional disc coverages (filling factors) of component $i$ and $F_i$ the corresponding disc-integrated brightness spectra.
The brightness spectra are time-independent but depend on wavelength and limb-angle.

The filling factors are derived from the reconstructed magnetic fluxes (Sect.~\ref{sect: mf_model}).
Sunspot filling factors (fractional surface areas) are taken directly from  observations (see Sect.~\ref{sect: Data}) and divided into umbra and penumbra with a fixed ratio of $\alpha_u / (\alpha_u + \alpha_p) = 0.2$, consistent with the mean observed value \citep{brandt_umbra-penumbra_1990, solanki_sunspots_2003}. 
As shown by \citet{chatzistergos_reconstructing_2021}, the precise umbra--penumbra partition has only a minute effect on the reconstruction.
Assuming a mean vertical magnetic field component of $\langle B_u\rangle = 1800\,\mathrm{G}$ in umbrae and $\langle B_p\rangle=550 \,\mathrm{G}$ in penumbrae \citep{keppens_magnetic_1996}, the total magnetic flux in sunspots is:
\begin{equation}
  \phi_\mathrm{s} = \phi_\mathrm{u} +\phi_\mathrm{p} = \alpha_\mathrm{u} \langle B_\mathrm{u}\rangle + \alpha_\mathrm{p} \langle B_\mathrm{p}\rangle. 
\end{equation}

Since ARs include sunspots and faculae, the facular flux is 
the remaining part of the total AR flux:
$\phi_\mathrm{f} = \phi_\mathrm{AR} - \phi_\mathrm{s}$.
The magnetic flux contributing to the network includes both the SSE flux and the open solar magnetic flux (OSF): $\phi_\mathrm{n} = \phi_\mathrm{SSE} + \phi_\mathrm{open}$. 
The filling factors for faculae and network are computed assuming a linear increase with mean flux density until a saturation threshold, above which they remain constant at unity:
\begin{equation}
\alpha_i = 
\min\!\left( \frac{B_i}{B_{\mathrm{sat},i}},\,1 \right),
\qquad 
i \in \{\mathrm{f},\mathrm{n}\},
\end{equation}
where $B_i = \phi_i/A_\odot$ and 
$A_\odot$ denotes the area of the solar hemisphere.
For the network we adopt $B_{\mathrm{sat,n}} = 800\,\mathrm{G}$, as estimated by \citet{krivova_reconstruction_2007,krivova_reconstruction_2010} from the analyses of magnetograms. 
The saturation threshold for faculae, $B_{\mathrm{sat,f}}$, is a free parameter determined from a comparison to observations, in the optimisation described below.
For the disc-averaged flux densities used here, the saturation limit is never reached. 
The quiet-Sun filling factor is then the remaining fraction of the solar surface not occupied by magnetic features, $\alpha_{\mathrm{q}} = 1 - \alpha_{\mathrm{u}} - \alpha_{\mathrm{p}} - \alpha_{\mathrm{f}} - \alpha_{\mathrm{n}}$.

The brightness spectra of the individual components
are unchanged compared to the previous SATIRE implementations. They
were computed by \citet{unruh_spectral_1999} using the ATLAS9 radiative transfer code \citep{kurucz_new_1993, kurucz_atlas12_2005}.
For the quiet Sun, penumbra, and umbra, we used Kurucz' model atmospheres with effective temperatures of 5777 K, 5400 K, and 4500 K, respectively.
The facular spectrum is based on the FAL-P atmosphere \citep{fontenla_energy_1993}, as modified by \citet{unruh_spectral_1999}.
These spectra represent the emergent intensities $I_i(\lambda,\mu)$, where $\mu$ is the cosine of the heliocentric angle, which we integrate over the visible disc
to obtain the disc-integrated component spectra $F_i(\lambda)$.

Faculae and sunspots are not evenly distributed over the disc but appear predominantly within the active-region belts: $\sim 5^\circ$--$30^\circ$ for spots and $\sim 5^\circ$--$45^\circ$ for faculae.
Therefore, a uniform disc integration would misrepresent their centre-to-limb behaviour.
Following \citet{krivova_reconstruction_2007,wu_solar_2018}, we integrate the intensity spectra
over the corresponding latitude belts and rescale the disc-integrated filling factors to the belt areas.

Spectra of the various components were computed under the assumption of local thermodynamic equilibrium (LTE), which becomes increasingly inaccurate below about 300~nm.
To remedy this, we applied an empirical correction following the same procedure as in earlier SATIRE versions \citep{krivova_reconstruction_2006,yeo_reconstruction_2014,chatzistergos_revisiting_2025}. It has been validated against non-LTE calculations by \citet{tagirov_readdressing_2019}.
Between 180 and 300~nm, the reconstructed SSI is adjusted to match the Whole Heliospheric Interval (WHI) reference spectrum \citep{woods_solar_2009}; between 115 and 180\,nm, it is linearly scaled to agree with
SORCE/SOLSTICE (The SOlar Stellar Irradiance Comparison Experiment on board of Solar Radiation and Climate Experiment) observations.  
These corrections are time-independent and therefore do not affect the long-term variability.

\subsection{Optimisation of model parameters}
\label{sect: optimisation}

Several free parameters enter the magnetic-flux evolution (Sect.~\ref{sect: mf_model}) and the irradiance computation (Sect.~\ref{sect: model irradiance}). 
Their values are constrained by comparison with three independent datasets (Sect.~\ref{sect: data optimisation}): the total photospheric magnetic flux \citep{arge_two_2002, wenzler_reconstruction_2006}, the OSF reconstruction of \cite{lockwood_reconstruction_2024}, and the TSI composite of \cite{montillet_data_2022}. 
We employ the genetic-algorithm code PIKAIA \citep{charbonneau_genetic_1995} to minimise the combined reduced $\chi^2$ between the model output and these observations. 
The resulting best-fit parameter set is listed in the bottom part of Table~\ref{tab: parameters}.
The top part lists the values of the fixed parameters, same for all runs.

\begin{table}[!thb]
\centering
\caption{Model parameters. }
\begin{tabular}{m{2cm}  m{1cm}  m{1cm}    m{1cm}   m{1cm}  } 
  \hline
  \hline Parameter                                     &\multicolumn{2}{c}{This study}&\multicolumn{2}{c}{\citet{krivova_modelling_2021}}\\
  &ISNv2 &     CEA17 &  ISNv2 & HoSc98   \\ 
    \hline
    $n_\mathrm{0} \, [\mathrm{cm^{-2}~day^{-1}}$]       & \multicolumn{4}{c}{$3.14 \times 10^{-14}$}\\[2pt]
    $\phi_\mathrm{0} \,[\mathrm{Mx}]$                  & \multicolumn{4}{c}{$10^{16}$}            \\ [2pt]
    $\phi_\mathrm{AR} \,[\mathrm{Mx}]$                 & \multicolumn{4}{c}{$4\times 10^{20}$}    \\ [2pt]
    $\phi_\mathrm{limit}\,[\mathrm{Mx}]$              & \multicolumn{4}{c}{$10^{23}$}             \\ [2pt]
    ${m}$                                             & \multicolumn{4}{c}  {-2.69}              \\ [2pt]
    ${\Delta m}$                                      & \multicolumn{4}{c}  {0.0946}             \\ [2pt]
    $\tau_\mathrm{AR}^\mathrm{0} \,[\mathrm{yr}]$     & \multicolumn{4}{c}  {0.027}              \\ [2pt]
    $\tau_\mathrm{SSE}^\mathrm{0} \, [\mathrm{yr}]$   & \multicolumn{4}{c}  {$1.1\times10^{-5}$}  \\ [2pt]
\hline
  \\
  $\phi_\mathrm{sat,f}\,[\mathrm{G}]$                           & 360.78 & 363.68 &       --        &       --    \\ [3pt]
  $\alpha$                                        & 0.07   &  0.09  &      0.059      &     0.075   \\ [3pt]
  $\tau_\mathrm{open}^\mathrm{r}\,[\mathrm{yr}]$  & 0.11   &  0.1   &       0.16      &     0.09    \\[3pt]
  $\tau_\mathrm{open}^\mathrm{s}\,[\mathrm{yr}]$   & 3.95   &  3.96  &       3.98      &       3.9   \\[3pt]
  $\tau_\mathrm{AR}^\mathrm{r}\,[\mathrm{yr}]$    & 2.68   &  2.94  &       2.79      &      1.57   \\ [3pt]
  $\tau_\mathrm{AR}^\mathrm{s}\,[\mathrm{yr}]$    & 57.0   & 42.4   &      88.15      &      89.81  \\[3pt]
  $\tau_\mathrm{SSE}^\mathrm{s}\,[\mathrm{yr}]$   & 10.21  & 11.26  &      10.15      &      10.11  \\[3pt]
  \hline
\end{tabular}\label{tab: parameters}
\tablefoot{Fixed parameters are listed in the top part, and those derived in the optimisation are in the bottom part. Columns 2–3 show values from this study; columns 4–5 show those from Krivova et al. (2021) for comparison. The second line of the header indicates the sunspot number record used as input.}
\end{table}

\begin{table}[ht]
\centering
\caption{Reduced $\chi^\mathrm{2}$ values between the model output and the comparison datasets. 
}
\begin{tabular}{m{2cm}  m{1cm}  m{1cm}    m{1cm}   m{1cm}  } 
  \hline
  \hline Parameter                                     &\multicolumn{2}{c}{This study}&\multicolumn{2}{c}{\citet{krivova_modelling_2021}}\\
  &ISNv2 &     CEA17 &  ISNv2 & HoSc98   \\ 
 \hline
 \vspace{0.1cm}
  $\mathrm{\chi^\mathrm{2}\,_{TF}}$   & 0.069 &    0.064 &  0.058   &  0.058  \\  
  $\mathrm{\chi^\mathrm{2}\,_{OSF}}$  & 0.321  &    0.322  &  0.389   &  0.219  \\
  $\mathrm{\chi^\mathrm{2}\,_{TSI}}$       & 0.074 &    0.079 &   --     &  --   \\
\hline \\
\end{tabular}\label{tab: RC_parameters}
\tablefoot{The comparison datasets are the total magnetic flux measured in ground-based magnetograms, the independent empirical reconstruction of the open solar magnetic flux, and the TSI measurement composite, see Sect.~\ref{sect: data optimisation} for data description.
Columns 1 and 2 list the results for the reconstructions using
ISNv2 and CEA17 sunspot number series as inputs. Columns 3 and 4 are the corresponding results from \citet{krivova_modelling_2021}. Note that their $\chi^2$ values of OSF refer to comparisons with the annual OSF reconstruction of \cite{lockwood_reconstruction_2014}}
\end{table}

\section{Results}\label{Sect: results}

\subsection{Magnetic flux}\label{sect: results mf}

The fractional disc coverages used in the irradiance reconstruction (Sect.~\ref{sect: model irradiance}) are derived from the modelled evolution of the solar surface magnetic flux. 
Because several aspects of the modelling differ from the earlier study of \cite{krivova_modelling_2021}, including updated input and reference datasets as well as the inclusion of TSI in the optimisation, we first compare our reconstructed total and open magnetic flux with the corresponding observations or independent reconstructions, as well as with the \cite{krivova_modelling_2021} version.
The free parameters of the model, optimised as described in Sect.~\ref{sect: optimisation}, are listed in the bottom part of Table~\ref{tab: parameters}.

\begin{SCfigure*}
            \includegraphics[width=0.66\textwidth]{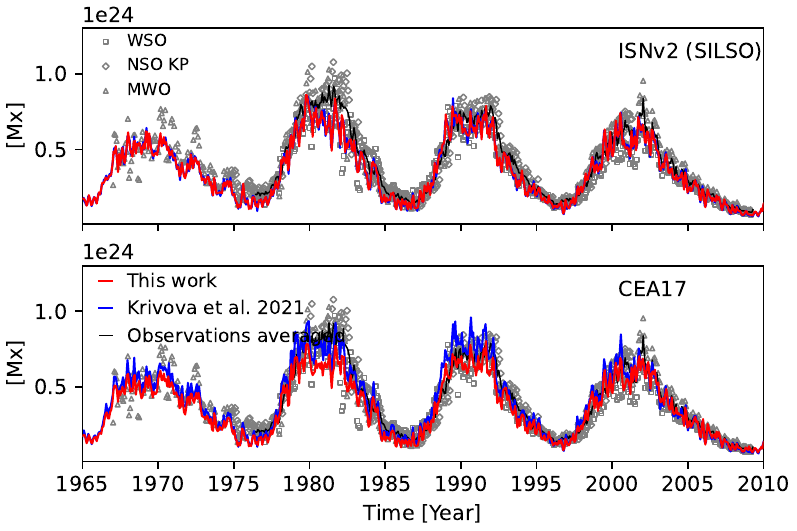}
            \vspace*{-4mm}
            \caption{Evolution of the total unsigned photospheric magnetic flux over time. 
            Red: our reconstructions based on the ISNv2 (top) and CEA17 GSN (bottom).
            Blue: reconstruction of \citet{krivova_modelling_2021} (ISNv2 in top panel, HoSc98 GSN in the bottom panel).
            Symbols: total magnetic flux  measurements from WSO, MWO, and NSO/KP; black line: their average.
            All quantities are shown as one-CR averages. The contribution of the SSE is scaled by a factor of 0.4 to compensate for the limited spatial resolution of observations.}
            \label{fig: total flux}
\end{SCfigure*}

To evaluate the reconstruction of the total photospheric magnetic flux, we compare the total unsigned flux with the combined measurements from WSO, MWO, and NSO/KP (Sect.~\ref{sect: data optimisation}), all at a CR cadence. 
Our two reconstructions~~-- ISNv2-based (top) and CEA17-based (bottom)~-- are shown in Fig.~\ref{fig: total flux} as 27-day (roughly one CR) means and are compared with the observational data, also at one-CR cadence.
\cite{krivova_effect_2004} showed that more than a half of the quiet-Sun flux remains undetected in NSO/KP synoptic charts due to the limited spatial resolution.
Therefore, following \cite{krivova_reconstruction_2007, krivova_modelling_2021}, we scaled the SSE contribution by a factor of 0.4 when comparing with observations.
We emphasise that this scaling was only applied for comparison purposes; when reconstructing the irradiance, all magnetic flux was retained.
Our tests showed that the exact value of this scaling factor has only a weak effect on the comparison presented in Fig.~\ref{fig: total flux}.
It mainly introduces a small overall shift in the absolute level but barely affects the variability, consistent with the fact that variations of the magnetic flux on time scales of a solar cycle and shorter on the Sun are dominated by ARs \citep{harvey_properties_1993,hofer_influence_2024}, see also the discussion of Fig.~\ref{fig: flux_contributions_isn} below.

Both our reconstructions agree very well with the observations. Comparing the 27-day average of our reconstruction to the averaged observations, we obtain a Pearson correlation coefficient of 0.95 and 0.94, a root mean square error (RMSE) of $8.51\times 10^{22} \mathrm{Mx}$ and $8.74\times 10^{22} \mathrm{Mx}$ and a reduced $\chi^\mathrm{2}$ value of 0.07 and 0.065 for ISNv2- and CEA17 GSN-based models, respectively (see Table~\ref{tab: RC_parameters}). 

Around the maximum of solar cycle 21, the agreement between our reconstructed total magnetic flux and the observation-average becomes somewhat poorer.
This period coincides with increased scatter among the magnetogram archives and with documented issues in the NSO/KP record before early 1992 (e.g., zero-point and calibration problems noted by \citealt{arge_two_2002}).
To account for these issues, \cite{wenzler_reconstruction_2006} applied a linear scaling to NSO/KP data over 1974--1987 to match the level of NSO data corrected by \cite{arge_two_2002}, by multiplying them by 1.242, which might not have been necessary.
When we instead use the NSO/KP values prior to that rescaling, the inter-archive scatter decreases, and the agreement with our reconstruction improves slightly (Fig.~\ref{fig: total flux_reduced nso}). 
Using this ``restored'' dataset, we obtain Pearson correlation coefficient of 0.95, an RMSE of $7.28\times 10^{22} \mathrm{Mx}$, and a reduced $\chi^2$ of 0.054 for the ISNv2-based reconstruction, while for the CEA-based reconstruction we get a Pearson correlation coefficient of 0.95, RMSE of $7.45\times 10^{22} \mathrm{Mx}$, and $\chi^2$ of 0.06.
This suggests that the residual (model--data) differences during cycle~21 are comparable to the archive-to-archive calibration uncertainties.

\begin{SCfigure*}
    \label{fig: flux_contributions_isn}
    \includegraphics[width=0.66\textwidth]{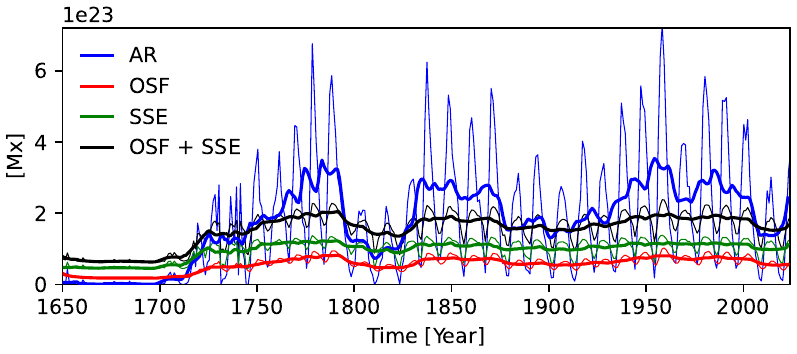}
    \vspace*{-4mm}
    \caption{Solar surface magnetic flux reconstruction based on ISNv2. Different components of  surface magnetic flux are shown with different colors, both as 11-year moving averages and annual averages shown as thick and thin lines, respectively. Shown are the contributions from: Active Regions (AR, blue), Open Solar Flux (OSF, red), Small-Scale Emergence regions (SSE, green), and the sum of OSF and SSE.}
\end{SCfigure*}

For comparison, Fig.~\ref{fig: total flux} also shows the total magnetic flux reconstruction by \citet{krivova_modelling_2021}. 
Our ISNv2-based model (top panel, red) is in excellent agreement with the reconstruction of \citet[][blue]{krivova_modelling_2021}.
The blue line is hardly visible, hidden behind the red curve.
Our CEA17-based reconstruction has lower amplitudes during solar cycles 21 and 22 than the one by \citet{krivova_modelling_2021} using the HoSc98 GSN. 
This difference is mainly due to the different GSN series used as input \citet{krivova_modelling_2021}. 
\cite{krivova_modelling_2021} reported $\chi^\mathrm{2}$ values of 0.058 for both of their reconstructions, ISNv2- and HoSc98-based. 
The slightly higher values obtained here can be attributed to differences in the input sunspot number records, the datasets used for model optimisation, as well as stricter constraints introduced by including TSI into optimisation. 

Figure~\ref{fig: flux_contributions_isn} shows the ISNv2-based reconstruction of the individual AR (blue) and small-scale (SSE+OSF; black) components of the magnetic flux (annual means and 11-year running means as thin and thick lines, respectively).
During the Maunder minimum ($\sim$1650--1700) and Dalton minimum ($\sim$1790--1830), most of the flux originates from SSEs, whereas AR flux is very low (Dalton minimum) or nearly absent (Maunder minimum).
The secular increase in the total flux is dominated by the AR component: the mean flux over 1967--2017 exceeds that over 1650--1700 by $\sim3.6\times10^{23}\,\mathrm{Mx}$, of which $\sim68$\% originates from ARs and 32\% 
from the combined SSE and OSF components. 
The CEA17-based reconstruction shows a very similar behaviour (Appendix~\ref{appendix_2}, Fig.~\ref{fig: flux_contributions_gsn}).

\begin{figure*}
 \includegraphics[width=1.0\textwidth]{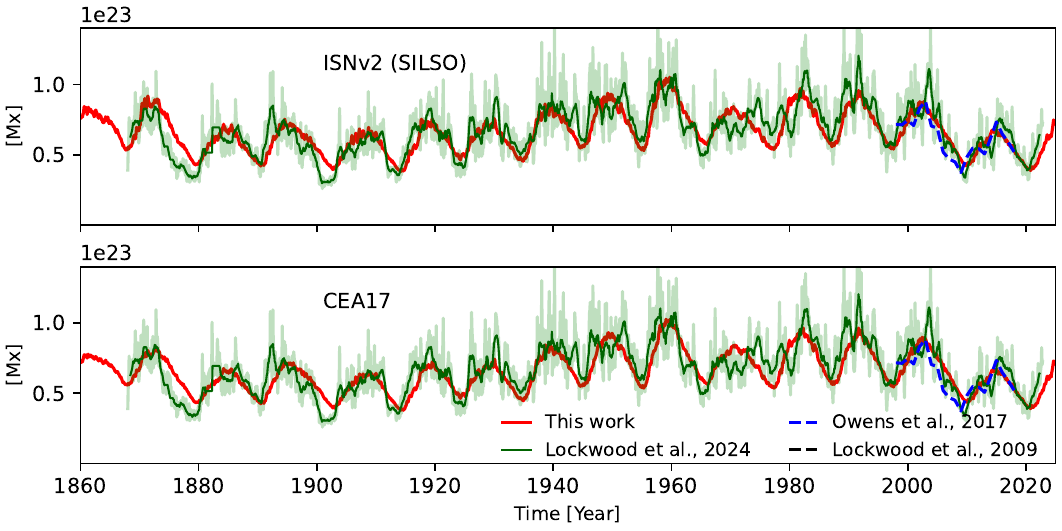}

    \caption{Evolution of the open magnetic flux over time. 
    Our reconstruction is shown in red for the ISNv2 input (top panel) and the CEA17 GSN (bottom panel), both at one-CR cadence. 
    In dark and light green, we show the OSF from \cite{lockwood_reconstruction_2024}, reconstructed from the geomagnetic aa-index for 13-CR and 1-CR cadence, respectively.
    The blue dashed line shows the in-situ OSF measurements from \cite{owens_sunward_2017} and the black dashed line shows the OSF estimates from \cite{lockwood_excess_2009}.}
    \label{fig: open flux}
\end{figure*}

Figure~\ref {fig: open flux} compares the evolution of the computed OSF to that reconstructed by \cite{lockwood_reconstruction_2024}.
The top panel presents the ISNv2-based reconstruction, and the bottom panel the CEA17-based reconstruction. For the optimisation of the model parameters (Sect.~\ref{sect: optimisation}), we used the OSF dataset at a cadence of one CR, available from 1868 to 2022. 

To quantify the performance of our model, we calculated reduced $\chi^2$ values relative to the one-CR OSF from \cite{lockwood_reconstruction_2024}.
Both of our reconstructions (ISNv2- and CEA17-based) show good agreement with these data. We obtain reduced $\chi^2$ values of 0.321 and 0.322 for our ISNv2 and CEA17 GSN-based reconstruction, respectively, when compared to the one-CR dataset. 
Comparing both our reconstructions with the 13-CR series, we get $\chi^2$ values of 0.228 and 0.225 for ISNv2 and CEA17 GSN reconstruction, respectively (Table~\ref{tab: RC_parameters}).
For the ISNv2-based model, the Pearson correlation with the 1-CR OSF is 0.7 and the RMSE of $1.44 \times 10^{22} \mathrm{Mx}$,
while for the CEA17-based model, these values are 0.7 and $1.45 \times 10^{22} \mathrm{Mx}$, respectively.

Our reconstructions also agree well with the directly measured heliospheric magnetic flux from \citet[][blue dashed line in Fig.~\ref{fig: open flux}]{owens_sunward_2017}, available at an annual resolution from 1998 to 2018.
Using annual means, we obtain for the ISNv2-based model a Pearson correlation coefficient of 0.87, a $\chi^2$ value of 0.313 and an RMSE of $0.66 \times 10^{22} \mathrm{Mx}$. For the CEA17 GSN-based model, the corresponding values are a Pearson correlation coefficient of 0.86, a $\chi$-square value of 0.35, and an RMSE of $0.72 \times 10^{22} \mathrm{Mx}$. For completeness, we also show the earlier part of the \citet{owens_sunward_2017} series, providing the OSF estimates derived from 
the OMNI interplanetary magnetic-field data for 1964--1998, corrected for solar-wind kinematic effects following \citet{lockwood_excess_2009}.  
These provide an observationally based but proxy-corrected estimate of OSF (black dashed line in Fig.~\ref{fig: open flux}).

\begin{SCfigure*}
    \includegraphics[width=0.66\textwidth]{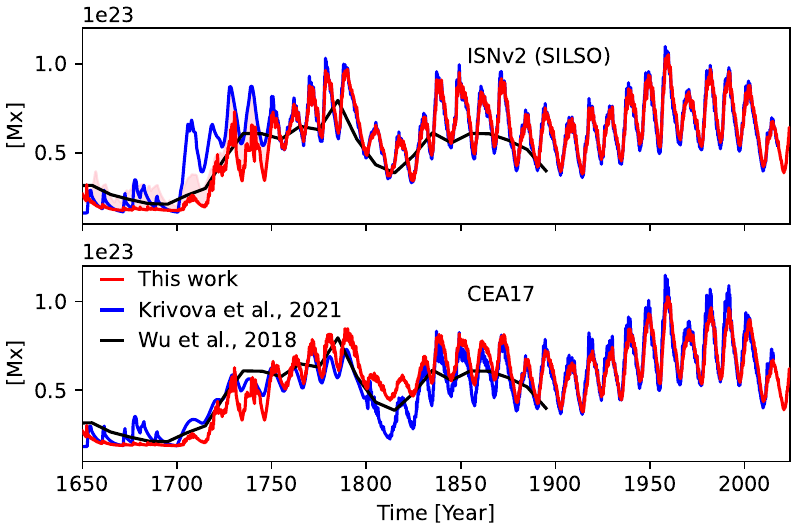}
\vspace*{-5mm}
    \caption{OSF reconstruction from this study (red) and by \citet[][blue]{krivova_modelling_2021} using ISNv2 (top)
    and GSN (bottom; CEA17 here and HoSc98 in \citealt[][]{krivova_modelling_2021}) as input.
    The pink shaded area shows our upper limit estimate for the Maunder Minimum, where the SN from \cite{carrasco_numerical_2025} was used as input.
    Black: OSF reconstructed from the cosmogenic isotope data by \cite{wu_solar_2018}.}
    \label{fig: open_flux_old}
\end{SCfigure*}

The OSF reconstruction back to 1650 is shown in Fig.~\ref{fig: open_flux_old}.
For 1750--2020, where the sunspot input is identical, our ISNv2-based OSF agrees well with that of \citet{krivova_modelling_2021}, with slightly lower cycle amplitudes due to the updated optimisation datasets. 
Before 1750, differences arise from the different sunspot records used: \citet{krivova_modelling_2021} combined annual ISNv2 and \citet{vaquero_level_2015}, whereas we employ daily HoSc98 GSN before 1749 and extend it further back with annual ISNv2 (to 1722) and the series of \citet{carrasco_understanding_2024}. 
The CEA17-based reconstruction likewise diverges from the HoSc98-based one before $\sim$1870, reflecting differences in the underlying GSN series.

We also compared our results with the OSF reconstructed from cosmogenic isotope data by \citet[][black curve]{wu_solar_2018}.
The OSF from \citet{wu_solar_2018} during the Maunder Minimum lies well within our upper and lower estimates. 
Since the revised model includes SSE emergence, some OSF is maintained even when sunspots are absent. 
For 1650--1670 we obtain mean OSF values of $0.182\times10^{23}\,\mathrm{Mx}$ (ISNv2) and $0.194\times10^{23}\,\mathrm{Mx}$ (CEA17). Given the scarcity and uncertainty of observations in this period, these should be interpreted as lower limits.

\subsection{Solar irradiance reconstruction}\label{sect: TSI}

We now present our irradiance reconstructions.
Model optimisation was performed using the TSI composite of \cite{montillet_data_2022}, applying an 81-day moving average. 
The 81-day smoothing applied to TSI reduces the influence of short-lived irradiance dips caused by individual large sunspot groups.
This ensures that the optimisation is driven primarily by variability on timescales of years and longer that the model is designed to reproduce.

Since the TSI composite of \citet{montillet_data_2022} differs in its absolute level from the TSIS-1/TIM record, we rescaled both the \citet{montillet_data_2022} composite and our reconstructed TSI to the TSIS-1/TIM level at the minimum of solar cycle 24.
The timing of this minimum (2019.96) was taken from the SILSO ISNv2 database\footnote{\url{https://www.sidc.be/SILSO/cyclesminmax}}.
For each series, we computed the mean of the daily TSI values within a 12-month window centred on 2019.96 (±6 months).
The ratio of the TSIS-1/TIM mean to the \citet{montillet_data_2022} mean defines the scaling factor applied to the composite.
A similar scaling was applied to the modelled daily TSI.

\begin{figure*}
  \centering\includegraphics[width=0.85\textwidth]{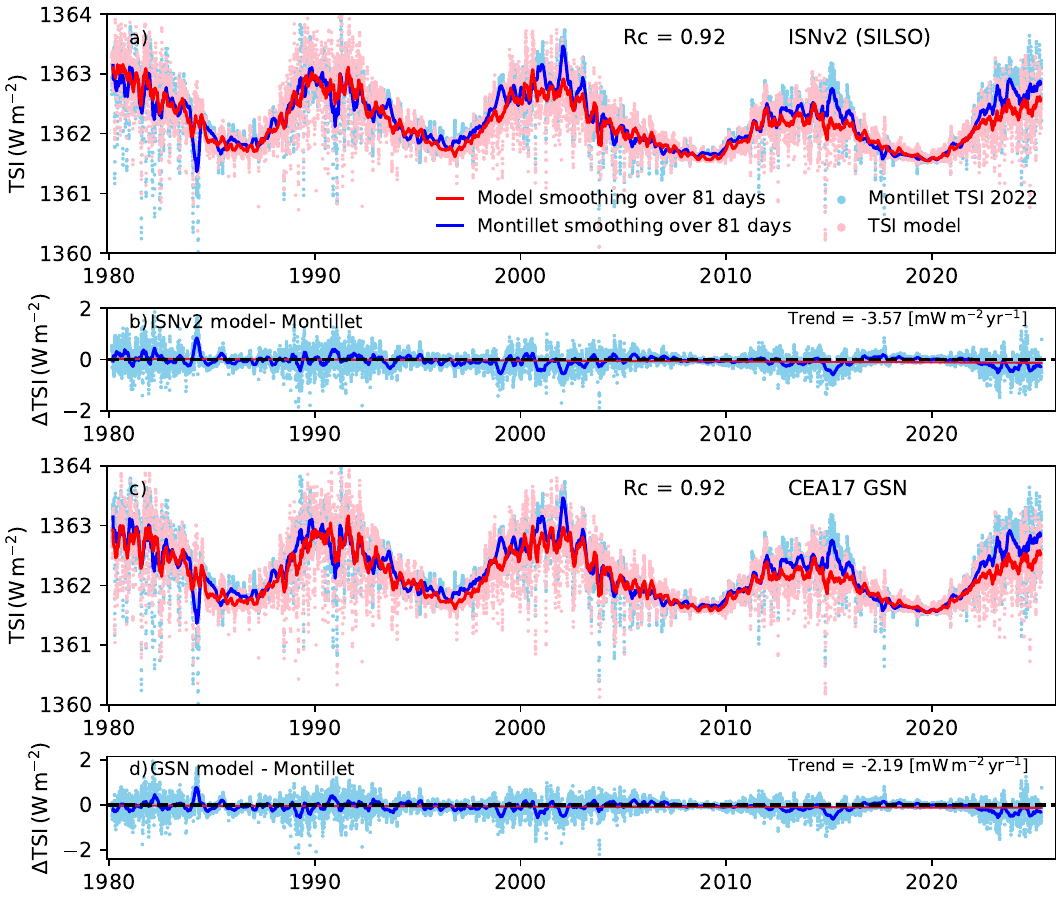}
\centering\includegraphics[width=0.85\textwidth]{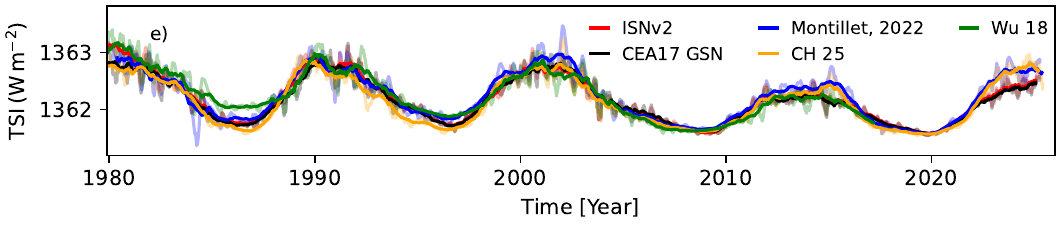}

    \caption{ a) Reconstructed TSI using ISNv2 as input (pink: daily values; red: 81-day smoothing).
    Blue dots and blue line: the TSI composite of \citet[][daily and 81-day smoothing, respectively]{montillet_data_2022}. 
    b) Difference between our reconstruction and the \cite{montillet_data_2022} composite (blue dots: daily; blue line: 81-day smoothing). 
    The horizontal black dashed line marks zero, and the orange line is the linear fit to the daily residuals.
      c) and d) same as a) and b), but using the CEA17 GSN as an input to the model. 
    e) Our TSI reconstructions based on ISNv2 and CEA17 (red and black). For comparison, the \citet[blue][]{montillet_data_2022} composite, the TSI reconstruction of \citet[green][]{wu_solar_2018}, and that of \citet[orange][]{chatzistergos_revisiting_2025} are shown. All datasets in panel e) are shown as 81-day (faint lines) and annual moving means (solid lines).}
    \label{fig: TSI_isn}
\end{figure*}

Figure~\ref{fig: TSI_isn}a shows our reconstructed TSI based on the ISNv2 in comparison with the TSI composite of \citet{montillet_data_2022}, both as daily values and their 81-day running means.
We find a good agreement between them, with Pearson correlation coefficients of 0.92 for the 81-day smoothed series and 0.80 for the daily values. The corresponding RMSE are $0.17\,\mathrm{W\,m^{-2}}$ for the 81-day series and $0.32\,\mathrm{W\,m^{-2}}$ for the daily one. 
The reduced $\chi^\mathrm{2}$ values for the 81-day smoothed comparison is $0.074$. 
Panel~(b) presents the residuals between our reconstruction and the \citet{montillet_data_2022} composite, daily (dots) and smoothed (line).
A linear fit to the daily residuals yields a trend of 
$-3.57\pm\,0.2\,\mathrm{mW\,m^{-2}\,yr^{-1}}$. This indicates that our reconstruction reproduces the 
long-term change of the satellite record without any significant secular bias.

Figure~\ref{fig: TSI_isn} panel~(c) shows the same comparison as panel (a) for our CEA17 GSN-based reconstruction. 
Similarly to the ISNv2-based case, we find a good agreement with observations, with Pearson's correlation coefficient of 0.92 for the 81-day smoothing and 0.8 for the daily values. The RMSE is 0.32~W\,m$^{-2}$ for the daily values and 0.18~W\,m$^{-2}$ for the 81-day smoothed ones. 
The reduced $\chi^2$ for the 81-day smoothing is 0.079.
A linear regression fit to the daily residuals gives a trend of $-2.19\,\mathrm{mW\,m^{-2} \,yr^{-1}}$ with uncertainty $\pm\,0.2\,\mathrm{mW\,m^{-2} \,yr^{-1}}$.

Further, we have compared the output of our model with the previous SATIRE-T version \citep{wu_solar_2018} and with SATIRE-S \citep{chatzistergos_revisiting_2025}, all shown in 
panel (e) of Fig.~\ref{fig: TSI_isn}. The \cite{wu_solar_2018} model is based on the previous SATIRE-T formulation and uses an earlier magnetic-flux evolution scheme by \citet{vieira_evolution_2010}.
The SATIRE-S model of \citet{chatzistergos_revisiting_2025}, in contrast, is the most accurate version of the SATIRE family, as it employs state-of-the-art full-disc magnetograms and continuum images. It is fully independent of sunspot numbers.

We also evaluated all these versions against the \citet{montillet_data_2022} composite over their overlap period (1980--2017), see table~\ref{tab: TSI_parameters_w_ch}.
Compared to the \citet{wu_solar_2018} reconstruction, our model gives comparable correlation coefficients (marginally lower/higher for daily/81-smoothed cadence), but a considerably lower $\chi^2$ and a significantly weaker trend in the daily residuals (see Table~\ref{tab: TSI_parameters_w_ch}), indicating an improved representation of long-term changes.
As expected, the SATIRE-S model \cite{chatzistergos_revisiting_2025} performs best overall.
Nevertheless, our model matches SATIRE-S remarkably closely,
despite relying solely on sunspot-number input.
This close agreement supports both the present model and the updated magnetic-flux evolution description of \citet{krivova_modelling_2021}.

\begin{table}[!htb]
\centering
\caption{Comparison of different TSI reconstructions with the TSI composite of \cite{montillet_data_2022}.}
\begin{tabular}{m{1.7cm} m{1cm} m{1cm} m{1cm} m{1.5cm}}
\hline
\hline
Model & $R_c$ daily & $R_c$ 81-day & $\chi^2$ & Trend [$\mathrm{mW\,m^{-2}\,yr^{-1}}$] \\
\hline
$\mathrm{ISNv2}$       & 0.79 & 0.91 & 0.079 & \,\   -4.39 \\
$\mathrm{CEA17,\,GSN}$ & 0.79 & 0.92 & 0.08  & \,\   -1.94 \\
$\mathrm{Wu18}$        & 0.79 & 0.89 & 0.12  &       -10.34 \\
$\mathrm{CH25}$        & 0.85 & 0.95 & 0.043 & \ \ \,\ 0.67 \\
\hline
\end{tabular}\label{tab: TSI_parameters_w_ch}
\tablefoot{Shown are the Pearson correlation coefficients for the daily and 81-day smoothed series, the chi-square values, and the trends in the daily residuals for our ISNv2 model, our CEA17 GSN model, the TSI reconstruction of \cite{wu_solar_2018}, and that of \cite{chatzistergos_revisiting_2025}. These values are calculated over the period 1980–2017, where data from all five datasets overlap.}
\end{table}

\begin{table}[!ht]
\centering
\caption{Pearson correlation coefficient between our TSI model based on ISNv2 and on CEA17 GSN in comparison to independent TSI datasets.}

\begin{tabular}{m{1.5cm}   m{1cm}    m{1cm}  m{1cm}  m{1cm} } 
  \hline
  \hline Model & VIRGO & TSIS TIM  &  DdW &  SORCE-TIM \\ \hline
  $\mathrm{TSI}_\mathrm{ISNv2}$        &    0.94 (0.79)  &   0.98 (0.85)  &    0.81 (0.69) &  0.86 (0.75)  \\  
  $\mathrm{TSI}_\mathrm{CEA17, GSN}$  &    0.94 (0.79) &     0.98 (0.84)  &    0.83 (0.71)  &  0.83 (0.74) \\  
\hline \\
\end{tabular}\label{tab: TSI rc parameters}
\tablefoot{The reconstructed TSI is compared to the VIRGO TSI measurements \citep{frohlich_virgo_1995,finsterle_total_2021}, TSIS-1/TIM \citep{pilewskie_tsis-1_2018}, \citet[][DdW]{dudok_de_wit_methodology_2017} composite, and SORCE/TIM \citep{kopp_total_2005}. Values refer to 81-day–smoothed series (daily in brackets).}
\end{table}

To evaluate the performance of our model more broadly, we compared our TSI reconstructions with several independent TSI datasets: the VIRGO TSI measurements \citep{frohlich_virgo_1995,finsterle_total_2021}, the composite of \citet{dudok_de_wit_methodology_2017}, TSIS-1/TIM \citep{pilewskie_tsis-1_2018}, and SORCE/TIM measurements \citep{kopp_total_2005}.
Pearson correlation coefficients, computed for both daily and 81-day–smoothed values, are summarised in Table~\ref{tab: TSI rc parameters}.
For the 81-day-smoothed series, the correlation with the various TSI data sets ranges between 0.81 and 0.98, while daily correlations range between 0.69 and 0.85. 
The best agreement is obtained with the TSIS-1/TIM and VIRGO measurements.

\begin{figure*}
    \includegraphics{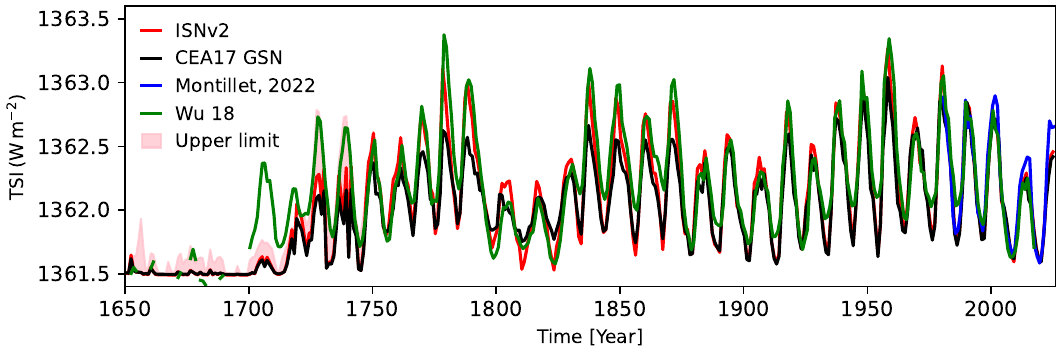}
\vspace*{-4mm}
\caption{Reconstructed TSI (annual means) since 1700 using ISNv2 (red) and CEA17 GSN (black) as input. The pink shaded area shows our upper limit estimate in TSI, where the \cite{carrasco_numerical_2025} SN were used.
            The TSI composite of \cite{montillet_data_2022} is shown in blue; 
            the TSI reconstruction of \cite{wu_solar_2018} in green. }
            \label{fig: TSI_full}
\end{figure*}

Figure~\ref{fig: TSI_full} presents the reconstructed TSI back to 1650 for both input series.
The ISNv2- and CEA17-based reconstructions are nearly identical before 1720 and after 1879, but differ between these periods, with the CEA17-based reconstruction showing weaker cycle amplitudes. Despite these differences, all three reconstructions (ISNv2-based, CEA17-based, and the upper-limit estimate) give a consistent TSI increase between the Maunder minimum and recent activity minima.

Comparing our reconstructed TSI during the Maunder Minimum (here defined as the 50-year mean over 1650--1700) with the recent period (1967--2017; 50-year mean), we find a TSI increase of $0.75\,\mathrm{W/\,m^2}$ for the lower-level estimate and $0.67\,\mathrm{W/\,m^2}$ for the upper-level estimate in the ISNv2-based reconstruction.
For the reconstruction using the CEA17 GSN series, we obtain an increase of $0.73\,\mathrm{W/\,m^2}$.
For comparison, \citet{wu_solar_2018} found an increase of $0.74\,\mathrm{W/\,m^2}$.
Overall, our reconstructions lie towards the lower end of published estimates of the TSI change since the Maunder Minimum, which reach up to $2.3\,\mathrm{W/\,m^2}$ \citep{penza_reconstruction_2024} and even $5.3\,\mathrm{W/\,m^2}$ \citep{egorova_revised_2018}.
For consistency, these values were computed from the respective published TSI series using the same definition adopted here, namely the difference between the 50-year mean over 1650--1700 and that over 1967--2017.
The inferred secular change remains below the upper limit of $2.0\pm0.7\,\mathrm{W/\,m^2}$ relative to the 2019 minimum
derived by \citet{yeo_dimmest_2020}, who constrained the minimum irradiance state of the Sun during a grand minimum using state-of-the-art magnetohydrodynamic simulations.

\begin{SCfigure*}
            \label{fig: TSI_contributions_isn}
    \includegraphics[width=0.66\textwidth]{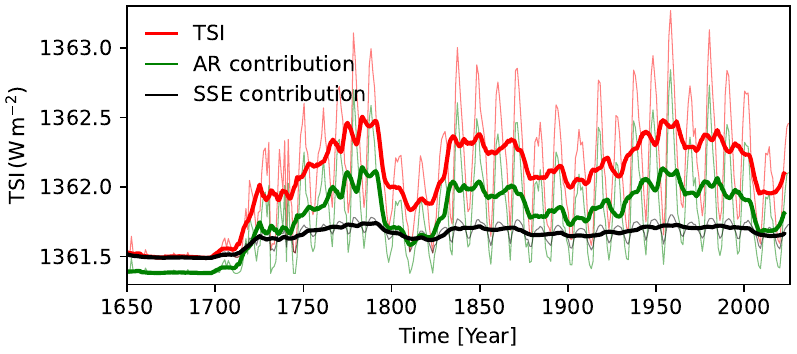}
\vspace*{-4mm}
            \caption{Total solar irradiance (TSI) reconstruction based on ISNv2 as an input (red). The green and black curves show the contribution of different components of magnetic flux to TSI. Green shows the contribution of AR magnetic flux to TSI, and black shows the contribution of small-scale emergences. Thin and thick lines show the annual means or 11-year running means, respectively.}
\end{SCfigure*}

Figure~\ref{fig: TSI_contributions_isn} illustrates how the individual magnetic components contribute to the reconstructed TSI in the ISNv2-based model.
Both ARs and the small-scale regions outside ARs (here, the sum of SSEs and the open flux) contribute to the secular change in TSI.
During the Maunder Minimum, when sunspots were essentially absent, the TSI variability originates mostly from the small-scale magnetic flux, while the AR contribution is negligible.
During the less deep and shorter Dalton minimum at the beginning of the 19th century, the two components are comparable.
Outside grand minimum conditions, ARs dominate nearly all solar cycle and multidecadal variability and provide the largest part of the long-term increase in TSI since the Maunder Minimum.
Quantitatively, from the Maunder minimum (average over 1650--1700) and the modern period (1967--2017), we obtain a TSI increase of $0.54\,\mathrm{W/m^2}$ from ARs flux and $0.21\,\mathrm{W/m^2}$ difference from the small-scale flux.
This means that $\sim72$\% of the secular rise originated from ARs and $\sim28$\% from the small-scale flux.

The same behaviour was found when using the CEA17 GSN record as input (see Appendix \ref{appendix_1}, Fig.~\ref{fig: TSI_contributions_gsn}). 
For that case, the TSI increase since the Maunder Minimum is $0.53\mathrm{W/m^2}$ ($\sim74$\%) for the AR component and $0.19\mathrm{W/m^2}$ $\sim26$\% from regions outside ARs.
Thus, while the small-scale component determines the baseline irradiance during extended periods of very low activity, the long-term rise in TSI since the Maunder Minimum was governed mainly by the evolution of active regions.

This contrasts with the irradiance modelling approach of \citet{penza_reconstruction_2024}, in which a substantial fraction of the secular variability is attributed to the network component through an empirically prescribed long-term modulation.
In our model, the secular variability arises self-consistently from the emergence, evolution, and decay of magnetic flux, including weak network and open flux contributions, without introducing an additional externally imposed long-term component.

\begin{SCfigure*}
\vspace*{-4mm}
            \includegraphics[width=0.66\textwidth]{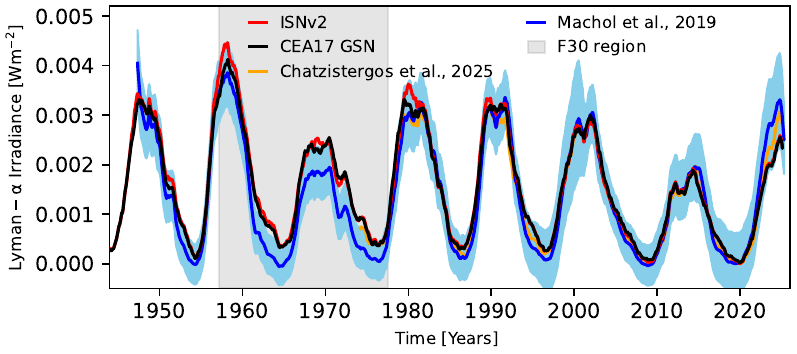}
            \caption{Reconstructed Lyman-$\alpha$ irradiance based on ISNv2 (red) and CEA17 (black) inputs.
            Also shown are the SATIRE-S Lyman-$\alpha$ reconstruction \citep[][orange]{chatzistergos_revisiting_2025}, and the observational composite by \citet[][blue curve]{machol_improved_2019} with its uncertainties shown in light blue. The grey shaded area marks the period when the F30 data were used by \cite{machol_improved_2019} instead of direct measurements. All series are shown as annual running means.}
            \label{fig: Lyman-alpha}
\end{SCfigure*}

The reconstruction of SSI is crucial for climate studies, particularly in the UV range, where irradiance variations strongly affect the chemistry of the Earth's atmosphere \citep{haigh_sun_2007,solanki_solar_2013}.
Of special importance is the Lyman-$\alpha$ line at 121.5~nm: it is the brightest emission line in the solar spectrum and varies by about a factor of two over the solar cycle \citep{woods_improved_2000,machol_improved_2019}.

Figure \ref{fig: Lyman-alpha} shows our Lyman-$\alpha$ reconstructions based on ISNv2 and CEA17 GSN, together with the observational composite of \citet[][]{machol_improved_2019}.
All series are displayed as annual running means.
Both reconstructions agree very well with the composite, with correlation coefficients of 0.924 (ISNv2) and 0.921 (CEA17 GSN) at daily cadence.
Importantly, no model parameters were explicitly adjusted to match the Lyman-$\alpha$ data.

We also show in Fig.~\ref{fig: Lyman-alpha} the SATIRE-S Lyman-$\alpha$ reconstruction of \citet[][]{chatzistergos_revisiting_2025}.
Despite being based on different model versions and on entirely independent input data, the two reconstructions agree closely. 
The Pearson correlation coefficients reach 0.92 for the daily ISNv2- and CEA17GSN-based models, respectively, and 0.99 for the annually smoothed series.

Both our (ISNv2- and CEA17 GSN-based) as well as the SATIRE-S \citep{chatzistergos_revisiting_2025} reconstructions, reproduce the Lyman-$\alpha$ composite well, except over 1960--1975.
During this interval (indicated as the shaded interval in the figure), no direct Lyman-$\alpha$ measurements were available, and the composite instead relies on values inferred from the 30~cm radio flux.
This suggests that the discrepancy possibly originates from uncertainties in the \citet[][]{machol_improved_2019} composite during this period rather than from our model.

\section{Conclusion}\label{conclusion}
We have presented new reconstructions of total and spectral solar irradiance over the past four centuries using the updated physics-based SATIRE-T model. 

In contrast to earlier SATIRE-T implementations, which relied on a more empirical parameterisation of small-scale magnetic flux emergence, the present reconstruction incorporates a revised description of the solar surface magnetic-field evolution, including a more realistic prescription for the emergence of small-scale magnetic flux linked to sunspot activity.
This provides a more self-consistent representation of irradiance variability, including its secular component.

Using two independent sunspot-number records (ISNv2 and CEA17 GSN) as input, we find that the resulting magnetic-flux and irradiance reconstructions are consistent with each other. 
Both versions reproduce the measured total photospheric magnetic flux from WSO, MWO, and NSO/KP, the open magnetic flux reconstructed from the geomagnetic aa-index \citep{lockwood_reconstruction_2024}, and the directly measured heliospheric magnetic flux from \citet{owens_sunward_2017}. 
The model also reproduces satellite TSI measurements (VIRGO, SORCE/TIM, TSIS-1/TIM, and the composite of \citealt{montillet_data_2022}) and the Lyman-$\alpha$ composite of \citet{machol_improved_2019} without any additional tuning.
The model achieves correlation coefficients of 0.81--0.98 with different space-based TSI records at 81-day resolution and 0.69--0.85 at daily cadence.
The correlation coefficient with the daily Ly-$\alpha$ composite is 0.92.

On centennial timescales, both sunspot inputs give a relatively small increase in TSI since the Maunder Minimum, between 0.67 and $0.75~\mathrm{W\,m^{-2}}$, depending on the sunspot record and in particular on the treatment of the data during the Maunder Minimum. 
In our model, the small-scale magnetic flux dominates the surface magnetic flux and the irradiance level during extended low-activity periods, while the main contribution (about 60--70\%) to the long-term rise from the Maunder Minimum to the present comes from active regions.

The close agreement between the two versions of the reconstruction, based on independent sunspot-number series, and with independent datasets, together with the more realistic treatment of small-scale magnetic emergence, strengthens confidence in the updated SATIRE-T irradiance series. 
These new reconstructions provide a physically consistent record of solar irradiance variability over the last 400 years and can be used in studies of solar influence on climate.

The SATIRE-T irradiance reconstructions produced in this work supersede earlier SATIRE-T versions and are available to the research community (see Data Availability). For the period after 1976, we recommend the SATIRE-S version by \citet{chatzistergos_revisiting_2025}, which is based on spatially resolved solar observations and therefore provides a more direct and accurate representation of the surface magnetic-field distribution.

\section*{Data availability}

All datasets used in this study are publicly available. 
The International Sunspot Number v2.0 can be obtained from the SILSO database 
(\url{https://www.sidc.be/SILSO/datafiles}). 
The CEA17 group sunspot number series and the updated sunspot-area record of \citet{mandal_sunspot_2020} are available at \url{https://www2.mps.mpg.de/projects/sun-climate/data.html}. 
The HoSc98 group sunspot number series is accessible at the National Geophysical Data Center. 
The annual sunspot numbers derived with the active-day fraction method are available from the authors of \citet{carrasco_relationship_2022,carrasco_understanding_2024,carrasco_numerical_2025}.

Magnetogram-based total magnetic flux data from WSO, MWO, and NSO/KP are publicly accessible through the respective observatories. 
The open‐flux reconstruction from the geomagnetic aa-index \citep{lockwood_reconstruction_2024} is available from \url{https://www.southampton.ac.uk/~gl/}. 
The in-situ heliospheric magnetic-flux data of \citet{owens_sunward_2017} and the OMNI‐based open-flux estimates corrected following \citet{lockwood_excess_2009} can be downloaded through NASA’s OMNI database (\url{https://omniweb.gsfc.nasa.gov/}).
The TSI composite of \citet{montillet_data_2022}, VIRGO TSI data, and the TSI composite of \citet{dudok_de_wit_methodology_2017} are available from PMOD/WRC. 
SORCE/TIM and TSIS-1/TIM TSI data are available from the LASP data repository. 
The Lyman-$\alpha$ composite of \citet{machol_improved_2019} is available from LASP. 
The SATIRE-S irradiance reconstruction of \citet{chatzistergos_revisiting_2025} is provided at \url{https://www2.mps.mpg.de/projects/sun-climate/data.html}.

The SATIRE-T irradiance reconstructions produced in this work 
are available under \url{https://www2.mps.mpg.de/projects/sun-climate/data.html}.

\begin{acknowledgements}
D.T. was supported by the International Max-Planck Research School (IMPRS) for Solar System Science at the Technical University of Braunschweig.
S. K. S and T.C. have received funding from the European Research Council (ERC) under the European Union's Horizon 2020 research and innovation program (grant agreement No. 101097844 — project WINSUN).
This research has made use of the Astrophysics Data System (ADS; \url{https://ui.adsabs.harvard. edu/}) Bibliographic Services, funded by NASA under Cooperative Agreement 80NSSC21M00561.
\end{acknowledgements}

\bibliographystyle{aa}
\bibliography{references}

\begin{appendix}
\section{NSO/KP calibration}
\label{appendix_1}

\begin{SCfigure*}
    \includegraphics[width=0.66\textwidth]{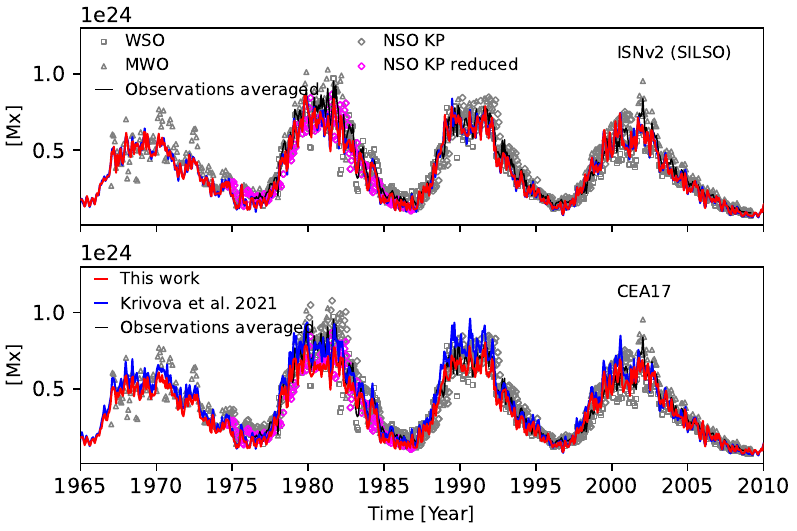}
    \caption{Same as Fig.~\ref{fig: total flux}, but here we reduced the NSO KP level in the period 1974--1987 to the level before the application of the \citet{wenzler_reconstruction_2006} calibration by dividing the 1974--1987 data by the factor 1.242.
    The black line shows the average over the magnetogram archives whenever at least two datasets are available.
    Using this ``restored'' NSO/KP dataset, we obtain Pearson correlation coefficients of 0.95 (ISNv2) and 0.95 (CEA17), with RMD values of -7.69\% and -7.75\%, respectively.
This suggests that the additional linear scaling applied by \citet{wenzler_reconstruction_2006} may not be appropriate for all magnetograms.
    }
    \label{fig: total flux_reduced nso}
\end{SCfigure*}

Figure~\ref {fig: total flux_reduced nso} illustrates the comparison between the total magnetic flux reconstructed by the model and measured in magnetograms, discussed in Sect.~\ref{sect: results mf}.
This figure is similar to Fig.~\ref{fig: total flux}, except that here NSO/KP data over 1974--1987 were re-scaled back to their original values from \citet{arge_two_2002} by dividing them by a factor 1.242 introduced by \citet{wenzler_reconstruction_2006}, see Sect.~\ref{sect: results mf}.
Restoring the NSO/KP data to the calibration of \citet{arge_two_2002} slightly improves the agreement with our model for both SN inputs. 
Using this ``restored'' dataset, we obtain
Pearson correlation coefficients of 0.95 (ISNv2) and 0.95 (CEA17), with RMSE of $7.28\times10^{22}\,\mathrm{Mx}$ and $7.45\times10^{22}\,\mathrm{Mx}$ 
and $\chi^2$ values of 0.054 and 0.06, respectively. This suggests that the additional linear scaling applied by \citet{wenzler_reconstruction_2006} may not be appropriate for all magnetograms.

\section{Contribution of different magnetic flux components to total solar magnetic flux and total solar irradiance}
\label{appendix_2}

\begin{SCfigure*}
    \includegraphics[width=0.66\textwidth]{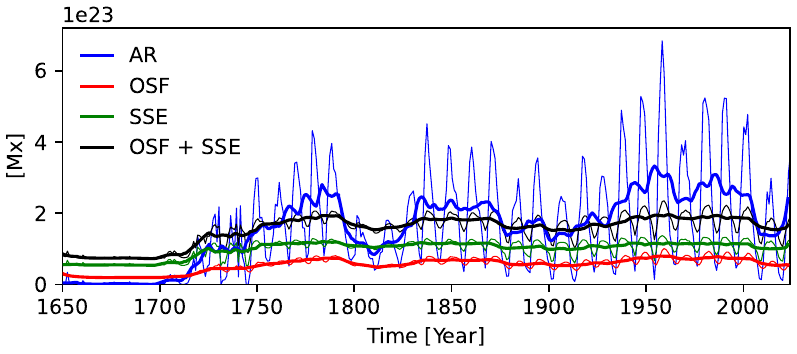}
    \caption{Same as Fig.~\ref{fig: flux_contributions_isn}, but using CEA17 GSN as an input.}
    \label{fig: flux_contributions_gsn}
\end{SCfigure*}

Figure~\ref{fig: flux_contributions_gsn} shows the CEA17-based reconstruction of the individual magnetic-flux components. As in the ISNv2-based case (Fig.~\ref{fig: flux_contributions_isn}), during the Maunder and Dalton minima, the surface flux is dominated by SSEs, with only partial (Dalton minimum) or negligible (Maunder minimum) AR contribution. The mean flux over 1967--2017 exceeds that over 1650--1700 by $3.6\times10^{23}\,\mathrm{Mx}$, with $\sim69$\% coming from ARs and 31\% from the sum of the SSE and OSF components.

\begin{SCfigure*}
            \centering
    \includegraphics[width=0.66\textwidth]{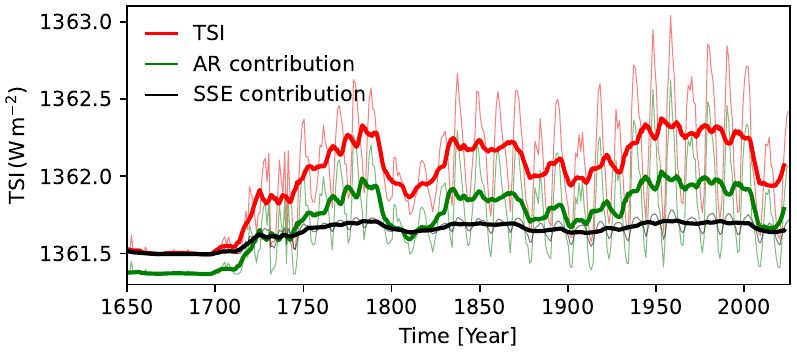}
            \caption{Same as Fig.\ref{fig: TSI_contributions_isn}, but here using the CEA17 GSN dataset as an input.}
            \label{fig: TSI_contributions_gsn}
\end{SCfigure*}

Figure~\ref{fig: TSI_contributions_gsn} shows the contribution of different magnetic flux components to the TSI reconstruction based on CEA17 GSN. We obtain similar results to those for the ISNv2-based reconstruction, where during periods of low activity (e.g., the Maunder minimum), the main contributor is the SSE flux, while for the rest of the time, the AR flux is the primary contributor to the TSI.

\end{appendix}

\end{document}